\documentclass[prb,twocolumn,showpacs]{revtex4}

\usepackage{amsmath,amsfonts,amssymb,bm}
\usepackage{dcolumn}
\usepackage[final]{graphicx}
\usepackage{bm}

\begin{document}

\bibliographystyle{apsrev}	

\title{Surface plasmons in doped topological insulators}

\author{Robert Sch\"utky}
\author{Christian Ertler}
\author{Andreas Tr\"ugler}
\author{Ulrich Hohenester}
\affiliation{Institut f\"ur Physik,
  Karl--Franzens--Universit\"at Graz, Universit\"atsplatz 5,
  8010 Graz, Austria}

\date{November 15, 2013}

\begin{abstract}
We investigate surface plasmons at a planar interface between a normal dielectric and a topological insulator, where the Fermi-energy lies inside the bulk gap of the topological insulator and gives rise to a two-dimensional charge distribution of free Dirac electrons.  We develop the methodology for the calculation of plasmon dispersions, using the framework of classical electrodynamics, with modified constituent equations due to Hall currents in the topological insulator, together with a Lindhard-type description for the two-dimensional charge distribution of free Dirac electrons.  For a system representative for Bi$_2$X$_3$ binary compounds, we find in agreement with recent related work that the modified constituent equations have practically no impact on the surface plasmon dispersion but lead to a rotation of the magnetic polarization of surface plasmons out of the interface plane.
\end{abstract}



\maketitle

\section{Introduction}\label{sec:introduction}

In the last couple of years the classification of matter states from a topological point of view has attracted a lot of interest.~\cite{hasan:10,qi:11}  Originally, this approach has been introduced to gain a better understanding  for the robustness of the Hall conductivity of the two-dimensional electron gas in the quantum Hall regime \cite{thouless:82}.  In contrast to quantum Hall systems, in topological insulators time reversal symmetry is fulfilled and they exhibit a strong spin orbit coupling leading to an inverted band structure.  Due to the bulk-boundary correspondence principle,~\cite{hasan:10,qi:11} three-dimensional topological insulators such as Bi$_2$Se$_3$, Bi$_2$Te$_3$ or Sb$_2$Te$_3$ show topologically protected surface states, rendering them robust against perturbations that are not breaking the time reversal symmetry.  In the long-wavelength limit the two-dimensional electron states at the surface can be described as massless Dirac electrons with the peculiar property that the spin is locked to the momentum, thereby forming a helical electron gas.  Thus, charge transport is closely intertwined with spin transport, which renders these systems interesting for spintronic applications.  In particular collective charge-density waves (plasmons) in a helical liquid are always accompanied by spin-density waves, so-called ``spin-plasmons'',~\cite{raghu:10} which might be utilized in spin accumulator devices.~\cite{appelbaum:11}

The surface states of the topological insulator reveal unique properties if an energy gap is induced in the dispersion by breaking the time reversal symmetry on the surface, e.g., by applying a perpendicular external magnetic field or by the proximity of magnetic materials.  Under this condition, a surface quantum Hall current with a half-integer quantized Hall conductivity arises, leading to an intriguing topological magneto-electric effect \cite{qi:11,essin:09}, in which an in-plane electric field induces an  antiparallel magnetization and analogously a charge polarization is induced by a magnetic field.  This effect can be accounted for by modified constituent equations, or more generally by a field theory including a $\theta \,\bm{E} \cdot\bm{B}$-term in the electrodynamic Lagrangian.  In systems with time-reversal symmetry, $\theta$ is quantized with $\theta=0$ or $\theta=\pi$ $({\rm mod}\,2\pi)$ for the topologically trivial and nontrivial case, respectively. 

As regarding the modified Maxwell equations, it is interesting to study electromagnetic properties at interfaces of media with abruptly changing $\theta$ values.  Shining polarized light onto a topological insulator leads to a small Kerr and Faraday effect, i.e., to a rotation of the light polarization when reflected or transmitted through a topological insulator film.~\cite{raghu:10,wang-kong:10,wang-kong:11}  However, the angles are not universal and depend on the optical constants of the involved media.  To separate the bulk contribution, a combined  measurement of the Kerr and Faraday angles has been proposed to find universal results depending only on the quantized $\theta$-value.~\cite{maciejko:10}  Interestingly, in thin layers of topological insulators a ``giant`` Kerr angle of $\pi/2$ has been predicted, due to the interference of multiple reflections as in a Fabry-Perot interferometer.~\cite{wang-kong:10,wang-kong:11}  Recent experiments confirm such large magneto-optical effects in thin layers of topological insulators.~\cite{aguilar:12}

Also the electromagnetic properties of surface plasmons in topolgical insulators become modified by the topological magneto-electric effect.  As shown by Karch~\cite{karch:11} the polarization of the surface plasmons is rotated from the transverse magnetic (TM) mode into the transverse electric (TE) direction by a small angle on the order of the fine structure constant $\alpha$.  In this work, the author has considered an interface between an insulator and a doped topological insulator, with the Fermi energy lying in the bulk conduction band, and the residual bulk charge density has been described as a Drude gas with negative permittivity $\varepsilon(\omega)$.  The Drude gas then allows to fulfill the usual condition for surface plasmon modes.~\cite{maier:07}
 
In this paper, we investigate surface plasmons in topological insulators in the case where the Fermi-energy lies \textit{inside} the bulk gap of the topological insulator, which means that the only free charges are Dirac-electrons on the surface, and the bulk contribution is completely described by its static dielectric constant (which is typically very large, around 40--80, for Bi$_2$X$_3$ binary compounds).  We develop the methodology for the classical electromagnetic fields of surface plasmons, with the quantum properties of the helical Dirac gas being included via its dielectric function, and compute the surface plasmon dispersion for realistic material systems.  In agreement with Karch,~\cite{karch:11} we find that the effect of the additional $\theta\,\bm E\cdot\bm B$ term in the Lagrangian has practically no influence on the surface plasmon dispersion, which is almost completely governed by the dielectric properties of the two-dimensional electron gas, but leads to a mixing of TE and TM field components.  This gives rise to a rotation of the magnetic polarization of surface plasmons out of the interface plane, which could be directly detected in experiment.

\section{Theory}\label{sec:theory}

In this section we derive the dispersion relation for surface plasmons at the planar interface between a topological insulator and a dielectric.  Our analysis closely follows the standard procedure for surface plasmons~\cite{maier:07} that has been recently extended to topological insulators.~\cite{karch:11}  However, contrary to the work of Karch,~\cite{karch:11} we treat the topological insulator as a true insulator that has a finite conductivity only at its surface.~\cite{hasan:10,qi:11} The structure under study thus consists of (i) the topological insulator (dielectric constant $\varepsilon_{2}$) for $z<0$, (ii) the surface of the topological insulator (two-dimensional conductivity $\sigma_{\rm 2D}$) at $z=0$, and (iii) a dielectric material (dielectric constant $\varepsilon_{1}$) for $z>0$.  The two-dimensional conductivity is computed for free electrons with a Dirac-like dispersion, similar to the case of graphene~\cite{wunsch:06,hwang:07} (for details see discussion below).  

\subsection{Surface plasmons}

Due to Hall currents at the surface of the topological insulator, the constituent equations for the electromagnetic response become~\cite{qi:11,karch:11}
\begin{align}\label{eq:constituent}
\bm D&=\varepsilon_0\varepsilon_r\bm{E}-\varepsilon_0\alpha\frac{\theta}{\pi}(c_0\bm{B})\nonumber\\
c_0\bm{H}&=\frac{c_0\bm{B}}{\mu_0\mu_r}+\alpha\frac{\theta}{\pi}\frac{\bm{E}}{\mu_0\mu_r}\,,
\end{align}
where $\bm{D}$ is the electric displacement, $\bm{E}$ the electric field, $\bm{B}$ the magnetic induction, and $\bm{H}$ the magnetic filed.  $\alpha$ is the fine-structure constant and the parameter $\theta$ is quantized and takes odd integer multiples of $\pi$ for topological non-trivial materials and zero otherwise (note that we consider for completeness the general case where the materials at both sides of the interface can be either dielectrics or topological insulators).  $\varepsilon_0$ is the permittivity of free space, $\varepsilon_r$ the dielectric function for the different materials, $\mu_0$ the permeability of free space, $\mu_r$ the relative permeability, and $c_0$ the vacuum speed of light.  Below we will use the abbreviations $\varepsilon=\varepsilon_0\varepsilon_r$, $\mu=\mu_0\mu_r$, and $c=\sqrt{\varepsilon_r\mu_r}\,c_0$.  

In what follows, we are seeking for eigenmodes that are confined to the interface.  We use a general wave ansatz for $\bm{E}$, which allows for mixed TE (transversal electric) and TM (transversal magnetic) modes
\begin{equation}\label{eq:modeansatz}
\bm E_i=\left[E_{\rm TE}\hat{\bm e}_y+\frac{c_i E_{\rm TM}^i}{\omega} 
  \left(k_x\hat{\bm e}_z-k_i^z\hat{\bm e}_x\right)\right]
  e^{i\left(k_x x+k_i^z z\right)}\,.
\end{equation}
The index $i$ distinguishes between the two different media. $E_{\rm TE}$ is the amplitude of the TE mode, $E_{\rm TM}$ the amplitude of the TM mode, and $\hat{\bm e}_x,\hat{\bm e}_y,\hat{\bm e}_z$ are the unit vectors that span our basis system. $c_i$, $\omega$, and $k$ are the speed, frequency, and wavenumber of light, respectively.  The relation between wavenumbers and frequency is governed by the usual dispersion relation $k_x^2+k_z^2=\varepsilon\mu\,\omega^2$.  A confined, plasmon-like mode is characterized by evanescent fields in the $z$ direction, i.e., $\Im m(k_1^z)>0$ and $\Im m(k_2^z)<0$ must hold.  The electric field of Eq.~\eqref{eq:modeansatz} and the magnetic induction are related through Faraday's law,~\cite{jackson:99} and the constituent equations [Eq.~\eqref{eq:constituent}] provide the connection to the dielectric displacement and magentic field.

To determine the unknown coefficients, we have to match the electromagnetic fields at the boundary.  The homogeneous Maxwell equations lead to the usual relations~\cite{jackson:99} $\bm{E}_{\| 2}=\bm{E}_{\| 1}$ and $B_{\bot 2}=B_{\bot 1}$, and the inhomogeneous equations yield
\begin{equation}\label{eq:boundary}
  \left(D_{\bot 2}-D_{\bot 1} \right)=\sigma\,,\quad
  \left(\bm{H}_{\| 2}-\bm{H}_{\| 1} \right)=\bm K\,,
\end{equation}
where $\sigma$ and $\bm K$ denote the surface charge and current distributions.  Ohm's law $\bm K=\sigma_{\rm 2D}\bm E_\|$ connects the currents at the surface of the doped topological insulator to the electric field, where $\sigma_{\rm 2D}$ is the two-dimensional optical conductivity, and the continuity equation provides the relation between $\bm K$ and $\sigma$.  For the electric field of Eq.~\eqref{eq:modeansatz}, we obtain
\begin{equation}
  j_y={\sigma}_{\rm 2D}E_{\rm TE}\,,\quad
  \sigma=-\frac{{\sigma}_{\rm 2D}c_1k_1^zk_x}{\omega^2}E_{\rm TM}^1\,.
\end{equation}
With these expressions, the boundary conditions of Eq.~\eqref{eq:boundary} become
\begin{subequations}
\begin{eqnarray}
  E_{\rm TM}^1&=&\left(\frac{\mu_1\left(\frac{k_2^z}{k_1^z}\right)-\mu_2}{\mu_1\mu_2}-
  \frac{{\sigma}_{\rm 2D} \omega}{k_1^z}\right)\frac{c_0\mu_0}%
  {\alpha \frac{\Delta\theta}{\pi} c_1}\,E_{\rm TE}\quad\\
  E_{\rm TM}^1&=&\frac{\alpha\left(\frac{\Delta\theta}{\pi}\right)\varepsilon_0c_0}%
  {c_1\left(\varepsilon_{1}-\varepsilon_{2}\frac{k_1^z}{k_2^z}\right)-
  \frac{{\sigma}_{\rm 2D}c_1k_1^z}{\omega}}\,E_{\rm TE}\,,\quad
\end{eqnarray}
\end{subequations}
with $\Delta\theta=\theta_1-\theta_2$.  To simultaneously satisfy the two equations relating the TE and TM amplitudes, we find the plasmon condition for topological insulators
\begin{eqnarray}\label{eq:plasmoncond}
  &&\left(\frac{\mu_1\left(\frac{k_2^z}{k_1^z}\right)-\mu_2}{\mu_1\mu_2}-\frac{{\sigma}_{\rm 2D} 
  \omega}{k_1^z}\right) \nonumber \\
  &&\quad\times\left(\varepsilon_{1}-\varepsilon_{2}\frac{k_1^z}{k_2^z}-
  \frac{{\sigma}_{\rm 2D}k_1^z}{\omega}\right)=
  \left(\alpha\frac{\Delta\theta}{\pi}\right)^2\frac{\varepsilon_0}{\mu_0}\,.\qquad
\end{eqnarray}
Note that for normal dielectrics with $\sigma_{\rm 2D}=0$ and $\Delta\theta=0$ this expression yields the usual surface plasmon conditions.~\cite{maier:07}  We can use Eq.~\eqref{eq:plasmoncond} to derive the dispersion relation for surface plasmons, which, however, cannot be expressed analytically, in contrast to metals~\cite{maier:07} or topological insulators without surface charges.~\cite{karch:11}  To obtain the dispersion, we (i) use the light dispersion $k_x^2+k_z^2=\varepsilon\mu\,\omega^2$ to express $k_{1,2}^z$ in terms of the parallel momentum $k_x$ and the light frequency $\omega$; for evanescent fields, the sign of $k_{1,2}^z$ has to be chosen such that the waves decay exponentially away from the interface.  For a given parallel wavenumber $k_x$, we then (ii) scan over frequencies $\omega$ in order to find those values where the plasmon condition of Eq.~\eqref{eq:plasmoncond} is fulfilled.  Within our computational approach this solution scheme is implemented within the software package Mathematica.

\subsection{Dielectric function of 2D electron gas}

The calculation of the dielectric function for a two-dimensional electron gas with a light-like dispersion closely follows Ref.~\onlinecite{wunsch:06,hwang:07}.  We first show how the optical conductivity $\sigma$ and the dielectric function are connected.  Quite generally, $\sigma$ can be related to the current-current correlation function,~\cite{kubo:85} which, within the random-phase approximation, only depends on the genuine electron-gas properties (without any Coulomb couplings).  However, to connect to the literature results~\cite{wunsch:06,hwang:07,jablan:09} we will first compute the dielectric function $\varepsilon$ for a two-dimensional electron gas embedded in a dielectric background with constant $\varepsilon_b$, and will then extract the optical conductivity from $\varepsilon$.  This procedure assumes a specific dielectric environment for the calculation of $\varepsilon$, but finally extracts the conductivity part which depends on the genuine electron gas properties only (all Coulomb coupling effects are properly included in our electrodynamic approach described above).

We start from Ohm's law $\bm J=\sigma\bm E$.  With the Fourier-transformed expressions for the continuity equation $\omega\rho=\bm q\cdot\bm J$ and the electric field $\bm E=-i\bm q V$, we can relate the induced charge distribution to the potential,
\begin{equation}
  \rho=\frac{\bm q\cdot\bm J}\omega=\frac\sigma\omega\,\bm q\cdot\bm E=
  -i\frac{\sigma q^2}\omega\,V\,.
\end{equation}
On the other hand, we can also use the dynamic polarization~\cite{wunsch:06} $P^{(1)}(\bm q,\omega)$ to relate these quantities through $\rho=P^{(1)}V$.  The dynamic polarization is related to the dielectric function according to $\varepsilon_r=\varepsilon_b-\nu(q)P^{(1)}(q,\omega)$, where $\nu(q)=1/(\varepsilon_0 q^2)$ and $\nu(q)=1/(2\varepsilon_0 q)$ are the Coulomb potentials for a three- or two-dimensional Fourier transform, respectively.  We thus find for the relation between the two-dimensional conductivity $\sigma_{\rm 2D}$ and the dielectric function the expression
\begin{equation}
  \sigma_{\rm 2D}(q,\omega)=-i\frac{2\varepsilon_0\omega}q
  \bigl[\varepsilon_r(q,\omega)-\varepsilon_b\bigr]\,.
\end{equation}

For the free, massless Dirac electrons confined to the geometry of the topological insulator, the Hamiltonian is of the form~\cite{qi:11,karch:11} $H=\hbar v_F\,\bm\sigma\cdot\bm k$, where $v_F$ is the Fermi velocity, $\bm\sigma=(\sigma_x,\sigma_y)^T$ a Pauli-vector matrix, and $\bm k$ the two-dimensional wavevector of the electrons.  The eigenstates $|\bm k,\pm\rangle$ with energy $E^\pm_{\bm k}=\pm\hbar v_Fk$ result from the diagonalization of the Hamiltonian, and can be associated with two cones above and below the Dirac point at $k=0$.~\cite{qi:11,karch:11}    Throughout we ignore effects of a ``mass term'' due to the symmetry-breaking magnetic field,~\cite{qi:11} which is responsible for the novel constituent equations of Eq.~\eqref{eq:constituent}.  This approximation is expected to hold for dopings where the chemical potential $\mu$ (measured with respect to the Dirac point) is much larger than the gap energy.  Below we will also need the overlap $f_{ss'}(\bm k,\bm q)=|\langle\bm k,s|\bm k+\bm q,s'\rangle|^2$ between two states, with $s=\pm$ labeling the upper and lower cone, which can be evaluated to
\begin{equation}
  f_{ss'}(\bm{k},\bm{q}) =\frac{1}{2} \left(1+ss'\frac{k+q \cos(\theta )}{\sqrt{k^2+2 k q \cos(\theta )+q^2}}\right)\,.
\end{equation}
Here $\theta$ is the angle between $\bm k$ and $\bm k+\bm q$.

\begin{figure}[t]
\centerline{\includegraphics[width=0.9\columnwidth]{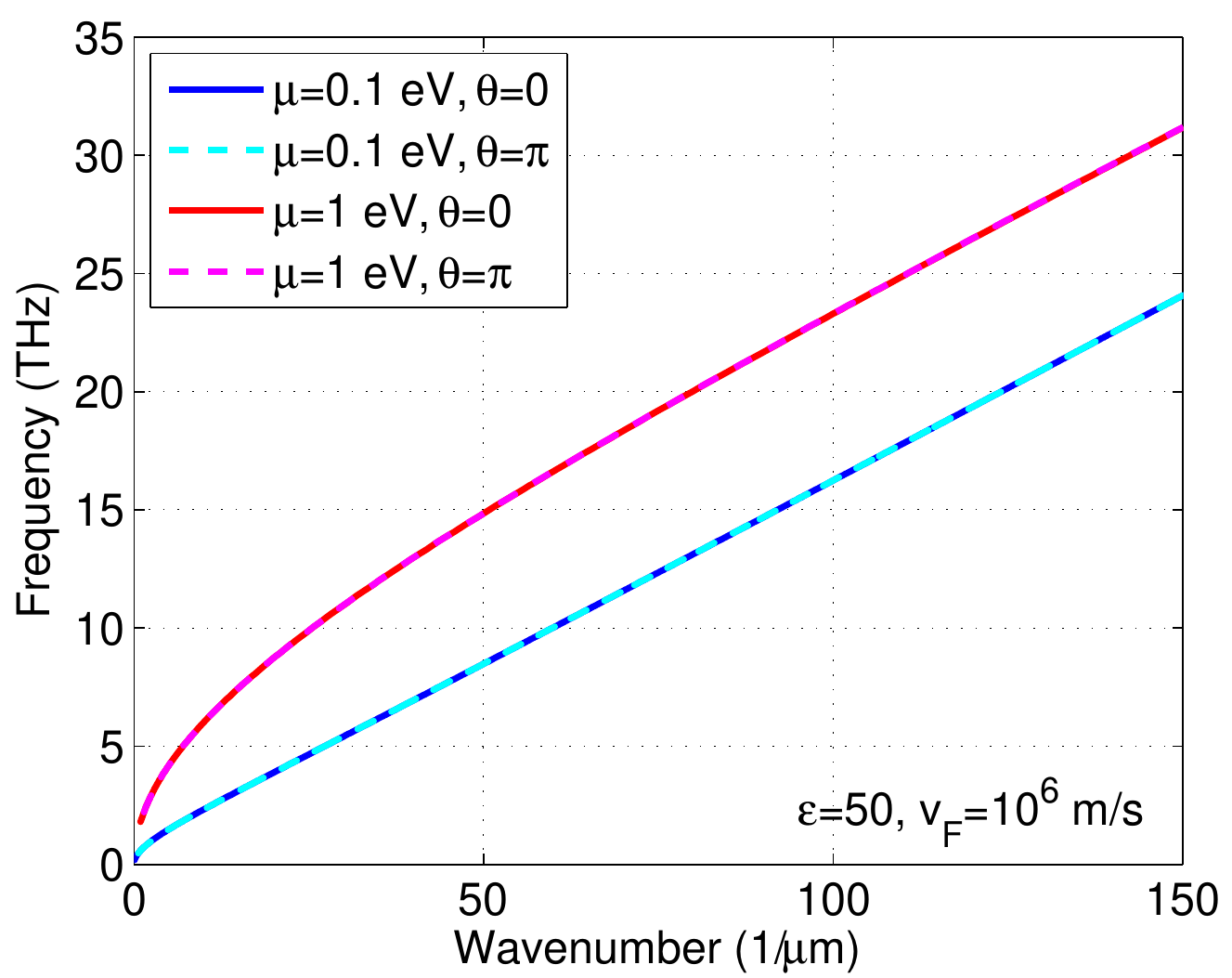}}
\caption{(Color online). Plasmon dispersion for a planar interface between a normal dielectric ($\varepsilon_{r1}=1$) and a topological insulator where the Fermi-energy lies inside the bulk gap of the topological insulator, leading to a two-dimensional charge distribution of free Dirac electrons at the surface.  We use material parameters representative for Bi$_2$X$_3$ binary compounds, namely a dielctric constant $\varepsilon_{r2}=50$ and a Fermi velocity $v_F=10^6$ ms$^{-1}$.  The plasmon dispersion is computed from Eq.~\eqref{eq:plasmoncond} for two doping levels of $\mu=0.1$ eV and $\mu=1$ eV, and the influence of the additional $\theta\,\bm E\cdot\bm B$ term in the Lagrangian is investigated by setting $\theta=\pi$ for the topological insulator and $\theta=0$ for a normal insulator.}
\end{figure}

The polarization within the random-phase approximation can be computed from Lindhard's equation~\cite{mahan:81,wunsch:06} 
\begin{eqnarray}\label{eq:lindhard}
  &&P^{(1)}(\bm{q},\omega ) = \frac{1}{4\pi ^2}\int d^2k \sum _{s,s'=\pm } 
  f_{ss'}'(\bm{k},\bm{q}) n_F\left(E^s_{\bm k}\right) \nonumber \\
  &&\times\left[\frac{1}{\hbar \omega +E^s_{\bm k}-E^{s'}_{\bm{k}+\bm{q}}+i0}-
  \frac{1}{\hbar \omega -E^s_{\bm k}+E^{s'}_{\bm{k}+\bm{q}}+i0}\right]\,.\nonumber\\
\end{eqnarray}
Here $n_F$ is the Fermi-Dirac distribution function and $i0$ is an infinitesimal quantity that ensures causality.  In the following we consider a doped topological insulator at zero temperature, where the electron states are filled up to the energy $\mu$ in the upper Dirac cone.  To evaluate the integrals, one conveniently sets $n_F=n_F^0+\delta n$, where $n_F^0$ and $\delta n$ denote the distributions for the undoped insulator and the doping contribution, respectively, which allows to convert the integrals into contributions for $n_F^0$ and $\delta n$ solely.~\cite{wunsch:06,schutky:13}  The Lindhard dielectric function of Eq.~\eqref{eq:lindhard} can then be solved analytically, and we obtain in accordance to Eqs.~(9--11) of Ref.~\onlinecite{wunsch:06} the final result (we set $g=1$)
\begin{eqnarray}
  &&P^{(1)}(q,\omega)=-\frac\mu{2\pi\hbar^2v_F^2}+\frac{F(q,\omega)}{\hbar^2 v_F^2}
  \Bigl\{G(x)-i\pi\nonumber\\
  && \qquad-\Theta(-x-1)[G(-x)-i\pi]-\Theta(x-1)G(x-1)\Bigr\}\,.\nonumber\\
\end{eqnarray}
Here $x=(\hbar\omega+2\mu)/(\hbar v_Fq)$, $\Theta$ is the Heaviside step function, and we have introduced the two complex functions $F(q,\omega)=(\hbar v_F^2q^2)/(16\pi\sqrt{\omega^2- v_F^2q^2})$ and $G(x)= x\sqrt{x^2-1}-\ln(x+\sqrt{x^2-1})$.

\section{Results}\label{sec:results}

We now apply the formalism developed in the previous section to a system with material parameters representative for Bi$_2$X$_3$ binary compounds, namely with dielctric constant $\varepsilon_{r2}=50$ and a Fermi velocity $v_F=10^6$ ms$^{-1}$.  For the medium in the upper half space we set $\varepsilon_{r1}=1$.  Figure 1 shows the plasmon dispersion computed from Eq.~\eqref{eq:plasmoncond} for two doping levels of $\mu=0.1$ eV and $\mu=1$ eV.  We investigate the influence of the additional $\theta\,\bm E\cdot\bm B$ term in the Lagrangian, by setting $\theta=\pi$ for the topological insulator and $\theta=0$ for a normal insulator.

\begin{figure}[t]
\centerline{\includegraphics[width=0.9\columnwidth]{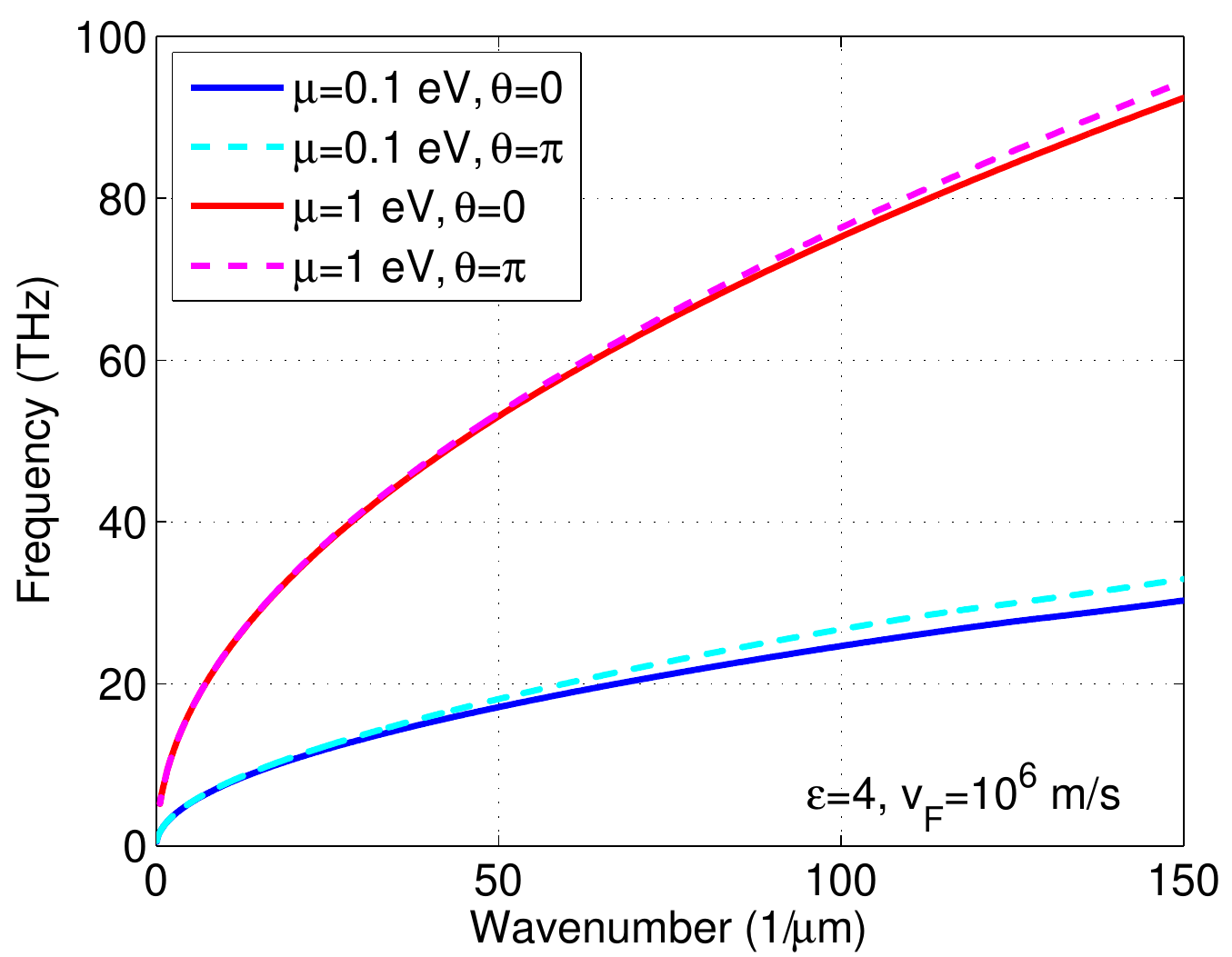}}
\caption{(Color online). Same as Fig.~1, but for a dielectric constant $\varepsilon_{r2}=4$ for the topological insulator.}
\end{figure}

As can be clearly seen from Fig.~1, the influence of the additional $\theta$-term on the dispersion relation of the surface plasmons is completely negligible.  The dispersion relation of surface plasmons is almost completely governed by the properties of the two-dimensional gas of Dirac electrons at the surface of the topological insulator (see Ref.~\onlinecite{pietro:13} for a recent experimental study).  Such plasmons have been studied in great detail for graphene, both theoretically~\cite{wunsch:06,hwang:07} and experimentally.~\cite{chen:12,fei:12}  As for graphene,~\cite{wunsch:06,hwang:07} the surface plasmon modes are not damped for sufficiently small frequencies, at least within the random-phase approximation under study.  The same is true for topological insulators and for the terahertz frequency regime.  Figure 2 shows the plasmon dispersion for a topological insulator where the background dielectric function has been artificially reduced to a value of $\varepsilon_{r2}=4$.  Some weak effect due to the $\theta$-term is visible at larger wavenumbers, where the dispersion curves with and without $\theta$-term start to deviate.  

We finally comment on the mode character of surface plasmons in graphene and topological insulators.  In graphene, surface plasmons are either purely TM or TE polarized, but become mixed when an external magnetic field is applied, resulting in very similar surface plasmon dispersions and mixing parameters as derived in this work.~\cite{gomez:12,iorsh:13}  In topological insulators the mixing between TM and TE components [see Eq.~\eqref{eq:modeansatz}] is governed by the modified constituent equations, see Eq.~\eqref{eq:constituent}.  For $\varepsilon_{r1}=1$ and $\varepsilon_{r2}=50$ we find on the vacuum side a ratio between TM and TE modes of $E_{\rm TM}^1:E_{\rm TE}\approx 1:2$, whereas for the reduced $\varepsilon_{r2}=4$ value we find modes with a strong TM character, $E_{\rm TM}^1:E_{\rm TE}\approx 10:1$.  

\section{Summary}

To summarize, we have investigated surface plasmons at a planar interface between a normal dielectric and a topological insulator, where the Fermi-energy lies inside the bulk gap of the topological insulator leading to a two-dimensional charge distribution of free Dirac electrons at the surface.  We have developed the methodology for the calculation of plasmon dispersions, using the framework of classical electrodynamics together with a Lindhard-type description for the two-dimensional charge distribution of free Dirac electrons.  The primary motivation for our study has been to investigate effects caused by the modified constituent equations, originating from an additional $\theta \,\bm{E} \cdot\bm{B}$-term in the electrodynamic Lagrangian, which leads to small Kerr and Faraday rotations.  For a system representative for Bi$_2$X$_3$ binary compounds, we have found that this additional $\theta$-term has practically no influence on the surface plasmon dispersion, which is almost completely governed by the dielectric properties of the two-dimensional electron gas, but leads to a mixing of TE and TM field components.  We have shown that modified dispersion relations might occur in other material systems, e.g., with strongly reduced background dielectric constants.  The methodology for surface plasmons in topological insulators with modified constituent equations due to surface Hall currents, which has been developed in this paper, is quite general and could be also used for the description of layer or multilayer systems.

\textit{Acknowledgement}.---This work has been supported by the Austrian Science Fund FWF under projects P24511 and P24248.

\end{document}